\documentclass[aps,showpacs,nofootinbib]{revtex4}

\usepackage{graphicx}

\newcommand{\be}{\begin{equation}}
\newcommand{\ee}{\end{equation}}
\newcommand{\ben}{\begin{eqnarray}}
\newcommand{\een}{\end{eqnarray}}

\newcommand{\la}{{\lambda}}

\newcommand{\cO}{{\cal O}}

\newcommand{\p}{\partial}
\newcommand{\na}{\nabla}

\newcommand{\ep}{\epsilon}

\newcommand{\ga}{\gamma}

\pacs{04.70.Bw}

\begin{document}

\title{Uniqueness theorem for stationary axisymmetric black holes in Einstein-Maxwell-axion-dilaton gravity}


\author{Marek Rogatko}
\affiliation{Institute of Physics \protect \\
Maria Curie-Sklodowska University \protect \\
20-031 Lublin, pl.~Marii Curie-Sklodowskiej 1, Poland \protect \\
rogat@kft.umcs.lublin.pl \protect \\
marek.rogatko@poczta.umcs.lublin.pl}

\date{\today}

\begin{abstract}
We prove the uniqueness theorem for stationary axisymmetric black holes solution
in Einstein-Maxwell-axion-dilaton gravity being the low-energy limit of the heterotic string
theory. We consider both the non-extremal and extremal Kerr-Sen black hole solutions.

\end{abstract}

\maketitle
\section{Introduction}
Much efforts have been devoted to the studies of the most striking investigations 
related to the black hole equilibrium states, the uniqueness theorem in
four-dimensional spacetime. The pioneering investigations 
were presented by Israel in Refs.\cite{isr67,isr68}. M\"uller zum Hagen {\it et al.} \cite{mil73} 
and Robinson \cite{rob77}  were able to find the generalization  of Israel's theorems.
It was shown that Schwarzschild and Reissner-Nordstr\"om (RN) solutions were the only
Einstein or Einstein-Maxwell (EM) (non-extreme) solutions that satisfied the 
conditions of being static  black hole metrics.
Quite different approach to the aforementioned problem using a conformal transformation on a spacelike hypersurface and 
the positive mass theorem \cite{sch79,wit81} 
was proposed in  \cite{bun87}-\cite{heu94}. 
The complete classification of four-dimensional vacuum black holes was established in Ref.\cite{chr99a,chr99b},
where the condition of non-degeneracy of the 
event horizon was removed.
Studies of the near-horizon geometry conducted in Ref.\cite{chr07}
enabled to eliminate the last restrictive condition of the static electro-vacuum
no-hair theorem.\\
The turning point for establishing the uniqueness of stationary
and axisymmetric black hole spacetimes being the solution of vacuum 
Einstein equations were achieved by Carter \cite{car73,car87} and
Robinson \cite{rob75}.
The systematic way of obtaining the desire results in
electromagnetic case was presented by Mazur 
\cite{maz82,maz01} and Bunting \cite{bun} (see also Ref.\cite{car85}). 
For a review of the uniqueness of black holes
solutions story in four-dimensions see, e.g, \cite{maz01} and \cite{heub}.
\par
Nowadays, there has been also an active period of constructing black hole solutions
in the string theories (see \cite{you97} and references therein). 
The stationary axisymmetric black hole solution in four-dimensional Einstein-Maxwell-axion-dilaton
(EMAD) gravity being the effective theory of the heterotic string theory was obtained by Sen \cite{sen}.
The $\sigma$-model representation and symmetries of stationary axisymmetric solutions in EMAD gravity
were widely studied \cite{gal94}-\cite{mat09}. One should also mention the efforts of proving the uniqueness theorem
for the black holes appearing in the theory in question \cite{mas93}-\cite{sim99}.
\par
Recently, the uniqueness theorem for the extremal black hole solutions acquired much attention.
It was shown that the near-horizon geometry of any extremal vacuum 
black hole must agree with the extremal 
Kerr black hole line element \cite{kun09}. The same statement is also valid in the case of Kerr-Newman spacetime.
The uniqueness for asymptotically flat stationary extremal vacuum black hole solutions in four and five-spacetime
dimensions were presented in Ref.\cite{fig10}. 
Using the Mazur identity it was proved that the only four-dimensional stationary axisymmetric asymptotically flat vacuum 
black hole with a single degenerate horizon was extremal Kerr solution or extremal Kerr-Newman in the case of 
EM theory \cite{ams10}.
The different method of proving the uniqueness theorem for extremal Kerr black hole was proposed in Ref.\cite{mei}
On the other hand, mathematically rigorous proof of the uniqueness theorem for the extremal Kerr-Newman 
black hole was elaborated in Ref.\cite{chr10}.
\par
In our paper we shall consider the stationary axisymmetric black hole solutions to the
EMAD gravity (Kerr-Sen solution) using $\sigma$-model representation of the underlying theory.
We shall take into account both the non-degenerate as well as the extremal Kerr-Sen black holes.
We prove the uniqueness theorem for both stationary axisymmetric black holes in this theory.
It enables us to close a gap that existed for many years in the proof
of the uniqueness theorem for black holes being the solutions of the low-energy effective heterotic string
theory.

Our paper is organized as follows. In Sec.II we review the $\sigma$-model
representation for the field equations
of EMAD gravity being the low-energy limit of the heterotic string theory.
This effective theory comprises gravitational field $g_{\mu \nu}$, scalar field dilaton,
the $U(1)$-gauge field and the Kalb-Ramond antisymmetric tensor field
which is equivalent in four-dimensional spacetime 
to the Peccei-Quinn antisymmetric pseudo-scalar axion. The next subsection will be devoted to the Mazur identity,
the main tool for proving the uniqueness theory for stationary axisymmetric black hole
in the theory in question.
In subsection C we determine the boundary conditions and show that two stationary
axisymmetric solutions of EMAD gravity subject to the same boundary and regularity
conditions coincide with each other. In Sec.III
the near-horizon geometry of the stationary axisymmetric
extremal solution to EMAD-gravity and the boundary conditions for the fields appearing
in the theory will be analyzed. All these help us to 
to find the uniqueness theorem for the extremal Kerr-Sen black hole solution.

\section{Uniqueness theorem for Kerr-Sen solution}
\subsection{$\sigma$-model representation for EMAD-gravity}

In this subsection we review the derivation of the three-dimensional $\sigma$-model
representation for the bosonic part of a heterotic string theory compactified to
four-dimensions. The so-called EMAD gravity contains metric tensor $g_{\mu \nu}$,
 $U(1)$ gauge field,
the Kalb-Ramond antisymmetric tensor $B_{\alpha \beta}$ and dilaton field $\phi$.
In four-dimensional spacetime the Kalb-Ramond tensor is equivalent to the pseudo-scalar
axion field. The resulting effective action for the bosonic sector of the heterotic string
with one gauge field is provided by
\be
S = \int d^4 x \sqrt{-g} \bigg(
R - 2 \na^{\mu} \phi \na_{\mu} \phi - {1 \over 2}~ e^{4 \phi}~ \na^{\mu} a \na_{\mu}
- e^{- 2 \phi}~F_{\mu \nu} F^{\mu \nu} - a F_{\mu \nu}~\ast F^{\mu \nu}
\bigg),
\ee
where $\ast F_{\mu \nu} = {1/2}~\ep_{\mu \nu \rho \ga} F^{\rho \ga}$.
In order to reduce the system under consideration to three-dimensions one should have a non-zero
Killing vector field. Introducing the timelike Killing vector field allows us to rewrite
a metric for an arbitrary stationary configuration in the form as \cite{isr72}
\be
ds^2 = - f~\bigg( dt - \omega_{a}~dx^{a} \bigg)^2 + {h_{ij} \over f}~dx^{i} dx^{j},
\label{wp}
\ee
where $h_{ij}$ is three-dimensional metric, $\omega_{a}$ is the {\it rotation} vector, while $f$ is scalar.
All the line coefficients depend on $x^{i}$, where $i = 1, \dots 3$.
\par
Due to the existence of timelike Killing vector field it will be possible to decompose
the generalized Maxwell equations into the two components field, {\it electric} and {\it magnetic}.
Namely, one has respectively the following relations \cite{gal94}-\cite{gal95b}:
\ben
F_{i0} &=& {1 \over \sqrt{2}} \p_{i} v,\\
e^{- 2 \phi}~F^{ab} &+& a~ \ast F^{ab} = {f \over \sqrt{2 h}}~\ep^{abc}~\p_{c} \kappa,
\een
where $v$ is the {\it electric} potential, while $\kappa$ is responsible for the {\it magnetic} one.
Further, one can introduce the {\it torsion} vector \cite{isr72} defined as
\be
\tau^{m} = - {f^2 \over \sqrt{h}} \ep^{mjk}~\p_{j}~\omega_{k}.
\ee
It turned out that it can be rewritten by means of the {\it twist} potential $\chi$
in the sense defined in Ref.\cite{isr72}
\be
\tau_{i} = \p_{i}\chi + v~\p_{i}\kappa - \kappa~\p_{i} v,
\ee
The most important feature of the above EMAD-gravity equations of motion is the fact that
all the above relations can be achieved by variation of the action provided by
\cite{gal96,mat09}
\ben
S &=& \int d^3 x \sqrt{h} \bigg( {}{}^{3} R +
{1 \over 2 f^2} \bigg( \p_{i}f~\p^{i}f + \tau_{i}\tau^{i} \bigg) - 2 \p_{m} \phi \p^{m} \phi
- {1 \over 2} e^{4 \phi}~\p_{m}a~\p^{m}a \\ \nonumber
&+& {1 \over f} \bigg(
e^{2 \phi} (\p_{i} \kappa - a \p_{i} v)(\p^{i} \kappa - a \p^{i} v) 
+ e^{- 2 \phi} \p_{k}v~\p^{k}v \bigg)
\bigg).
\een
Consequently,
it can be proved that the above equations can be cast into a set of relations derived from the action
for $\sigma$-model for {\it vector potential} $\Phi_{A}$ 
coupled to three-dimensional gravity.
Namely, the action is provided by \cite{gal95}
\be
S = \int d^3 x \sqrt{h} \bigg( 
{}{}^{(3)}R - G_{AB}~\p_{i}\Phi^{A}~\p_{j} \Phi^{B}~h^{ij}
\bigg),
\label{sig}
\ee
where the line element of the target space implies
\ben
G_{AB}~d\Phi^{A}~d\Phi^{B} &=& 
{1 \over 2} e^{2(\eta - \zeta)}~da^2 + d\eta^2 + d\zeta^2 - e^{2 \zeta} dv^2 -
e^{2 \eta} (d \kappa - a dv)^2 \\ \nonumber
&+& {1 \over 2}e^{2(\eta - \zeta)}(d \chi + v~d\kappa - \kappa~dv)^2,
\een
where we have denoted $\eta = \phi - {1/2} \ln f$ and $\zeta = - (\phi + 1/2 \ln f)$.
The {\it vector potential} $\Phi_{A}$ has six non-zero components $(f,~v,~\kappa,~\chi,~a, ~\phi)$.
From the action (\ref{sig}) we can get the standard equations of motion for gravitating $\sigma$-model.
They imply
\be
{}{}^{(3)}R = G_{AB}~\p_{i}\Phi^{A}~\p_{j} \Phi^{B}~h^{ij}.
\ee
Varying the action (\ref{sig}) with respect to $\Phi_{A}$ we arrive at  the equation of motion for
$\Phi^{A}$ field
\be
{}{}^{(h)}\na_{i}~{}{}^{(h)}\na^{i} \Phi^{A} + 
h^{ij}~\Gamma^{A}{}{}_{BC}~\p_{i}\Phi^{B}~\p_{j} \Phi^{C} = 0,
\ee
where ${}{}^{(h)}\na_{i}$ is the nabla operator with respect to three-dimensional
metric $h_{ij}$, while $\Gamma^{A}{}{}_{BC}$ is the Christoffel symbol in the target space metric
$G_{AB}$.
\par
Now, we proceed to consider another Killing vector field which is assumed to commute with
the timelike one introduced before. It implies,
without loss of generality, that the three-dimensional metric can be expressed as follows:
\be
dh^2 = e^{2 \ga} (d \rho^2 + dz^2) + \rho^2~d\phi^2,
\ee
Hence, the equation of motion for $\Phi_{A}$ field yields
\be
{}{}^{(h)}\na_{i}~{}{}^{(h)}\na^{i}~\Phi^{A} + \Gamma^{A}{}{}_{BC}~
\bigg( \p_{\rho}\Phi^{B}~\p_{\rho}\Phi^{C} + \p_{z}\Phi^{B}~\p_{z}\Phi^{C}
\bigg) = 0.
\ee
On this account, the function $\ga$ may be determined by the equations
\ben
{}{}^{(h)}R_{\rho \rho} &-& {}{}^{(h)}R_{z z} = {2~ \p_{r} \ga \over \rho} = 
G_{AB}~\bigg( 
 \p_{\rho}\Phi^{A}~\p_{\rho}\Phi^{B} + \p_{z}\Phi^{A}~\p_{z}\Phi^{B}
\bigg),\\ 
{}{}^{(h)}R_{\rho z} &=& {\p_{z} \ga \over \rho} = G_{AB}~\p_{\rho}\Phi^{A}~\p_{z}\Phi^{B}.
\een
Then, the action of the system can be rewritten in terms of the {\it current} matrix $J^{i}$ 
\cite{gal96,yur01,yur01a}
\be
S = {1 \over 4} \int d\rho~dz~\rho~ Tr (J^{i} J_{i}),
\ee
where $J^{i} = \na^{i} M ~M^{-1}$. Using the Gauss decomposition, the symmetric matrix $M$ may be written
as
\be
M = \pmatrix{ P^{-1} &  P^{-1}~Q \cr
Q~ P^{-1} & P + Q~ P^{-1}~Q \cr},
\ee
while symmetric two-dimensional matrices $P$ and $Q$ are given by the following:
\be
P = \pmatrix{ f - e^{-2 \phi}~v^2 & - e^{-2 \phi}~v \cr
- e^{-2 \phi}~v & - e^{-2 \phi} \cr},
\qquad
Q = \pmatrix{ - \chi + v~w & w \cr
w & - a \cr},
\ee
where we have denoted $w = \kappa - a~v$.
\subsection{Mazur identity}
In the proof of the uniqueness theorem for stationary axisymmetric black hole solution in
EMAD gravity a key role will be played by the so-called Mazur identity \cite{maz82,heub}.
Let us consider two sets of field configurations $M_{[0]}$ and $M_{[1]}$ satisfying
the equation of motion of the underlying theory and denote
the difference $J_{diff}$ between those two field configurations by the following
relation:
\be
J^{i}_{diff} = J^{i}_{[1]} - J^{i}_{[0]} = \na^{i} M_{[1]}~M_{[1]}^{- 1}
- \na^{i} M_{[0]}~M_{[0]}^{- 1}.
\ee
Then, we define the {\it deviation} matrix 
$\Psi$ which implies
\be
\Psi = M_{diff}~M_{[0]}^{- 1} = M_{[1]}~M_{[0]}^{- 1} - {\bf 1},
\label{dev}
\ee
where ${\bf 1}$ is the unit matrix and $M_{diff} = M_{[1]} - M_{[0]}$.
One can remark that the {\it deviation} matrix $\Psi$ will be equal to zero matrix
if and only if the two field configurations will accord. 
Expressing the Laplacian of the {\it deviation} matrix in the manner of $\rho$ and $z$-coordinates
one arrives at the following:
\be
\p_{i} \bigg(
\rho~\p^{i} tr \Psi \bigg) = \rho~h_{ij}~\bigg(
J^{t}_{i~diff}~M_{[0]}^{- 1}~J^{i}_{diff}~M_{[1]} \bigg),
\ee
where we define the transpose current matrix in the form as follows:
\be
J^{t}_{i} = M^{- 1}~\na_{i} M~.
\ee
By integrating the above equation over the adequate region $\Omega$ of the $(\rho,~z)$-plane
and by means of the Green theorem we get the expression
\be
\int_{\p \Omega} \rho~\p^{m} (tr \Psi)~dS_{m} =
\int_{\Omega}~ \rho~h_{ij}~\bigg(
J^{t}_{i~diff}~M_{[0]}^{- 1}~J^{i}_{diff}~M_{[1]} \bigg)~d\rho~dz.
\ee
The boundary of the region in question embraces the black hole event horizon, plane of rotation
and the infinity.\\
Because of the fact that matrix $M$ has a square root matrix $m$,
i.e., $M = m~m^{t}$ one can rewrite the Mazur identity in a suitable form.
On this account, one arrives at
\be
\int_{\p \Omega} \rho~\p^{m}( tr \Psi)~dS_{m} =
\int_{\Omega}~ \rho~h_{ij}~\bigg(
tr \Phi^{t}_{i}~\Phi^{i} \bigg)~d\rho~dz,
\label{mm}
\ee
where $\Phi_{k} = m_{[1]}~J_{k~diff}^{t}~m_{[0]}^{- 1}$.
It can be seen that the right-hand side of Eq.(\ref{mm}) is non-negative. If one impose the
boundary conditions on $\p \Omega$ such that the left-hand side of the relation
disappears,
we can conclude that $J_{diff}^{i} = 0$. From the relation 
$\na^{i} \Psi = M_{[0]}^{- 1}~J^{i}_{diff}~M_{[1]}$ it also follows that the matrix $\Psi$ has to be
constant over the considered region $\Omega$. In particular,
if the matrix in question is equal to zero matrix, then it yields that two solutions
$M_{[0]}$ and $M_{[1]}$,
subject to the same boundary and regularity conditions match each other.

\subsection{Boundary conditions}
In this subsection we shall apply the Mazur identity for the field configuration described
by the {\it current} matrix $J_{i}$.
We shall look for the boundary conditions of the fields at infinity,
on the plane of rotation as well as at the black hole event horizon. Moreover,
we assume asymptotical flatness of the considered black hole solution to EMAD gravity
and regularity on a plane of rotation and on the the black hole event horizon.
\par
The Kerr-Sen metric being the stationary axisymmetric solution to EMAD gravity was
derived in Ref.\cite{sen}. The metric of the Kerr-Sen black hole spacetime implies
\be
ds^2 = - {\Delta \over \Sigma}
\bigg( d{\hat t} - a \sin^2 \theta~d {\hat \phi} \bigg)^2
+ {\sin^2 \theta \over \Sigma}\bigg[
\bigg( {\hat r} ({\hat r} - r_{-}) + a^2 \bigg)~d {\hat \phi} - a d {\hat t}
\bigg]^2 + {\Sigma \over \Delta}~d {\hat r}^2 + \Sigma~ d \theta^2,
\ee
where we have defined the following quantities:
\be
\Sigma = {\hat r}~ ({\hat r} - r_{-}) + a^2~\cos^2 \theta, \qquad 
\Delta = ({\hat r} - r_{-})~({\hat r} - 2 M) + a^2, \qquad r_{-} = {Q^2 \over M}.
\ee
$M$ is the black hole mass, $Q$ is attributed to its charge, while $a$ is 
the Kerr rotation parameter related to the black hole
angular momentum by $J = a~M$. The black hole event horizon radii are located at
\be
{\hat r}_{\pm} = M + {r_{-} \over 2} \pm \sqrt{\bigg(
M - {r_{-} \over 2} \bigg)^2 - a^2}.
\ee
Furthermore, the asymptotic behaviour of the metric coefficients are provided by relations
\ben \label{gg} \label{met}
g_{{\hat t}{\hat t}} &=& - \bigg( 1 - {2 M \over {\hat r}} + \cO \bigg({1 \over {\hat r}^2}\bigg) \bigg), \qquad
g_{{\hat t}{\hat \phi}} = - {2 J \sin^2 \theta \over {\hat r}} + \cO \bigg({1 \over {\hat r}^2} \bigg),\\ \nonumber
g_{{\hat \phi} {\hat \phi}} &=& \bigg [ {\hat r} ( {\hat r} - r_{-}) + \cO ({\hat r}) \bigg] \sin^2 \theta, \qquad
g_{\theta \theta} = {\hat r} ( {\hat r} - r_{-} ) + \cO ({\hat r}),
\label{gg1}
\een
where $M$ and $J$ are respectively mass and angular momentum
asymptotically conserved at spherical spatial infinity.\\
For the fields existing in the theory under consideration,
i.e., for dilaton field, {\it electric} and {\it magnetic} part of $U(1)$ gauge field and for the axion one,
we have the following limits as ${\hat r}$ tends to infinity:
\be
\phi \simeq \cO \bigg({1 \over {\hat r}} \bigg), \qquad v \simeq \cO \bigg({1 \over {\hat r}} \bigg),
\qquad \kappa \simeq \cO \bigg({1 \over {\hat r}^2} \bigg), \qquad a \simeq \cO \bigg({1 \over {\hat r}^2} \bigg).
\label{limit}
\ee
On the other hand, the fields existing in in the theory under consideration, 
i.e., dilaton and axion fields and {\it electric} and {\it magnetic} part of $U(1)$ gauge field and for the axion one,
yield
\ben \label{fields}
e^{2 \phi} &=& {({\hat r} - r_{-})^2 + a_{k}^2~\cos^2 \theta \over \Sigma}, \qquad
a = {r_{-}~a_{k}~\cos \theta \over ({\hat r} - r_{-})^2 + a_{k}^2 ~\cos^2 \theta},\\
v &=& (2 M r_{-})^{-{1 \over 2}}~{({\hat r} - r_{-}) \Sigma}, \qquad
\kappa = {r_{-}~ a_{k}~\cos \theta \over \Sigma}.
\label{fields1}
\een
They have the following limits as $r$ tends to infinity:
\be
\phi \simeq \cO \bigg({1 \over {\hat r}} \bigg), \qquad v \simeq \cO \bigg({1 \over {\hat r}} \bigg),
\qquad \kappa \simeq \cO \bigg({1 \over {\hat r}^2} \bigg), \qquad a \simeq \cO \bigg({1 \over {\hat r}^2} \bigg).
\ee
As in Ref.\cite{mor04},
comparing the asymptotic forms of the metric tensor coefficients given by relations
(\ref{met}) with the Weyl-Papapetrou metric (\ref{wp}), we derive the boundary conditions.
On the other hand, one can assume that they behave in some manner 
and conduct the uniqueness proof \cite{mor08}.\\
Moreover, the regularity conditions on the rotation plain requires that $g_{\phi \phi}$ should have the form
$g_{\phi \phi} = f_{\phi \phi}~\sin^2 \theta$. It implies that $f_{\phi \phi} \simeq
{\hat r} ({\hat r} - r_{-}) + \cO({\hat r})$.
\par
The event horizon has $S^2$-topology which enables one to introduce
the spheroidal coordinates
\be
z = \la~ \mu, \qquad \rho^2 = (\la^2 - c^2)(1 - \mu^2),
\ee
where $\mu = \cos 2 \theta$.
The boundary $\la = c$ is responsible for the black hole event horizon, while
two rotation axis segments distinguishing the north and south of the horizon
are given by the respective limit
$\mu = \pm 1$.
Consequently, the asymptotic behaviour of $f_{\phi \phi}$ implies
\ben
\mu &\rightarrow& \pm 1~~\Rightarrow f_{\phi \phi} \simeq \cO(1), \qquad
\la \rightarrow c~~ \Rightarrow f_{\phi \phi} \simeq \cO(1), \\ \nonumber
\la &\rightarrow& \infty~~\Rightarrow  f_{\phi \phi} \simeq
\la + {r_{-} \over 2}~\sqrt{\la} - {r_{-}^2 \over 4}
+ \cO \bigg( \la^{- {1 \over 2}} \bigg).
\een
Similarly, one has that 
\be
g_{\phi \phi} = {\rho^2 \over f} - f \omega^2 = 
\bigg [ {\hat r} ({\hat  r} - r_{-}) + \cO ( {\hat r}) \bigg] \sin^2 \theta,
\ee
where $f = f_{\phi \phi}\sin^2 \theta$. It may be noted that
$\rho$ is provided by the following relation:
\be
\rho^2 = \bigg[ {\hat r}^2 ({\hat r} - r_{-})^2 + \cO ( {\hat r}) \bigg] \sin^2 \theta.
\ee
By virtue of the above,
it can be seen that $\rho$ vanishes at the $\phi$-invariant plane, where $\sin \theta = 0$,
and also on the event horizon due to the form of the metric (\ref{wp}).
It is also clear from the definition of $\la$, that
the relation between $\la$ and ${\hat r}$ yields
\be
\la = {\hat r} ({\hat r} - r_{-}) + \cO \bigg({1 \over {\hat r}} \bigg),
\ee
while for $r$ as a function of $\la$, we get
\be
{\hat r} = \sqrt{\la} \bigg( 1 - {r_{-} \over 8 \la} + \cO \bigg({1 \over {\hat r}}\bigg) \bigg)
+ {r_{-} \over 2}.
\ee
It follows directly that,
the asymptotic behaviours of $\rho$ are determined by the formulae
\ben \label{as1}
\mu &\rightarrow& 1~~ \Rightarrow   \rho \simeq \cO \bigg( \sqrt{1 - \mu} \bigg),\\ \label{as2}
\mu &\rightarrow& - 1~~ \Rightarrow  \rho \simeq \cO \bigg( \sqrt{1 + \mu} \bigg),\\ 
\la &\rightarrow& c~~ \Rightarrow \rho \simeq \cO \bigg( \sqrt{\la - c} \bigg),\\
\la &\rightarrow& \infty~~\Rightarrow  \rho \simeq \cO \bigg( \la \bigg),
\label{aass3}                     
\een
On the other hand, the following form of equation may be noted for the {\it rotation} vector:
\be
\omega_{\phi} = - {2 J \over {\hat r}^2 ({\hat r} - r_{-})} + \cO \bigg({\hat  r}^{-5} \bigg).
\ee
\par
As usual we take into account
the domain of outer communication $<<{\cal J}>>$ 
as an oriented  rectangle. Namely, one has the following:
\ben
\p {\cal J}^{(1)} &=& \{ \mu = 1,~ \la = c, \dots, R \},\\ \nonumber
\p {\cal J}^{(2)} &=& \{ \la = c,~ \mu = 1, \dots, -1 \},\\ \nonumber
\p {\cal J}^{(3)} &=& \{ \mu = - 1,~ \la = c, \dots, R \},\\ \nonumber
\p {\cal J}^{(4)} &=& \{ \la = R,~ \mu = - 1, \dots, 1 \}.
\een
The corresponding metric on the domain of outer communication $<<{\cal J}>>$, written in spheroidal coordinates
is given by
\be
ds_{<<{\cal J}>>}^2 = (\la^2 - c^2 \mu^2) \bigg(
{d \la^2 \over \la^2 - c^2} + {d \mu^2 \over 1 - \mu^2} \bigg).
\ee
Then, the boundary integral on the left-hand side of the Mazur identity (\ref{mm})
may be written as
\be
\int_{\p {\cal J}} \rho~\p^{a} (tr \Psi)~dS_{a} =
\int_{\p {\cal J}} d\la~ \rho~\sqrt{{h_{\la \la} \over h_{\mu \mu}}} \p_{\mu} (tr \Psi)
- 
\int_{\p {\cal J}} d\mu ~ \rho~\sqrt{{h_{\mu \mu} \over h_{\la \la}}} \p_{\la} (tr \Psi).
\label{bound}
\ee
We can also readily write down
the asymptotic forms of $\sqrt{{h_{\la \la} / h_{\mu \mu}}}$ and
$\sqrt{{h_{\mu \mu} / h_{\la \la}}}$. They become respectively
\be
\mu \rightarrow 1~~\Rightarrow \sqrt{{h_{\la \la} \over h_{\mu \mu}}} \rightarrow \cO \bigg( \sqrt{1 - \mu}\bigg),
\qquad
\mu \rightarrow - 1~~\Rightarrow \sqrt{{h_{\la \la} \over h_{\mu \mu}}} \rightarrow \cO \bigg( \sqrt{1 + \mu}\bigg),
\ee
and similarly we achieve
\be
\la \rightarrow c~~\Rightarrow \sqrt{{h_{\mu \mu} \over h_{\la \la}}} \rightarrow \cO \bigg( \sqrt{\la - c}\bigg),
\qquad
\la \rightarrow \infty~~\Rightarrow \sqrt{{h_{\mu \mu} \over h_{\la \la}}} \rightarrow \cO \bigg( \la \bigg).
\ee
In what follows we have used MAPLE symbolic mathematical program
to calculate the trace of {\it deviation} matrix.
\par
First let us consider the behaviour of the {\it deviation} matrix near axes. From
Eqs.(\ref{as1}) and (\ref{as2}) we see that $\rho \rightarrow \cO(\sqrt{1 - \mu})$ when $\mu$ tends to $1$,
or $\rho \rightarrow \cO(\sqrt{1 + \mu})$ when $\mu \rightarrow -1$. The same kind of behaviour reveals the coefficients
on the right-hand side of the integral of the Mazur identity Eq.(\ref{bound}). On the other hand,
when we approach the axes of rotation $f_{\phi \phi} \simeq \cO(1)$. Similarly,
inspection of Eqs.(\ref{fields})-(\ref{fields1}) reveals that $v,~a,~e^{2 \phi}$ and $\kappa$
tend to $\cO(1)$ as we approach the considered limit.
\par
It remains to take into account the behaviour of the {\it deviation} matrix near the black hole event horizon.
From the previous considerations one has that if $\la \rightarrow c$
$f_{\phi \phi} \simeq \cO(1)$ and $\rho \rightarrow \cO(\sqrt{\la - c })$. The inspection of the
behaviour of the fields in EMAD-gravity given by the relations
(\ref{fields}) and (\ref{fields1})
enables to conclude that as $\la \rightarrow c$
they are proportional to  $\cO(1)$. They are well-behaved functions as $\la$ tends to the constant value $c$.
The same behaviours reveal the coefficients in relation (\ref{bound}). Summing it all up,
we draw the conclusion that it is sufficient to establish, that $tr~\Psi \simeq \cO(1)$ near the event horizon of
Kerr-Sen black hole.
\par
On the other hand, at spatial infinity $f_{\phi \phi} \simeq \cO(\la)$ and the coefficients
in the Mazur integral given by Eq.(\ref{bound}) are of the same form. The same tendency is provided by
$\rho$. Then, the careful inspection of all the components building the trace of the {\it deviation} matrix $\Psi$
reveals that $\p_{\mu} (tr \Psi) \simeq \cO (\la^{-4})$ plus terms of order $\la^n$, where $n > -4$.
\par
Just we have shown that $tr \Psi$ is bounded in the orbit space and vanishing at spatial infinity.
It concludes that two configurations conditions coincide, i.e., $M_{[0]} = M_{[1]}$ in all the
domain of outer communication $<<{\cal J}>>$. It provides that two solutions of equations of motion for
EMAD-gravity underlying the same boundary and regularity conditions coincide with each other for at least one point in
$<<{\cal J}>>$. We can assert to the coclusion.\\
{\bf Theorem}:\\
Let us consider a stationary axisymmetric solution to four-dimensional EMAD-gravity being the low-energy
limit of the heterotic string theory with asymptotically timelike Killing vector field $k_{\mu}$
and spacelike Killing vector field $\phi_{\mu}$ responsible for rotation. Then, 
any solution with the same boundary and regularity conditions as the Kerr-Sen black hole is the Kerr-Sen
solution itself.

\section{Uniqueness of extremal Kerr-Sen solution}
Consider now, the extremal Kerr-Sen solution, i.e., ${\hat r}_{+} = {\hat r}_{-}$. Our next task will be to
extract the near-horizon geometry of the extremal solution. To proceed further, let us define the coordinate
transformation in the form as follows:
\ben
{\hat t} &=& 2M~{(M - {r_{-} \over 2}) \over (M + {r_{-} \over 2})~\la}~t, \\ 
{\hat \phi} &=& \phi + {(M - {r_{-} \over 2}) \over (M + {r_{-} \over 2})~\la}~{\hat t},\\
{\hat r} &=& {\la \over y}~\bigg(M + {r_{-} \over 2} \bigg) + \bigg(M + {r_{-} \over 2} \bigg).
\een
Taking the limit of the scaling parameter $\la \rightarrow 0$, we arrive at the near-horizon geometry
of the Kerr-Sen extremal black hole spacetime. Consequently, we obtain the metric which yields
\ben \label{nh}
ds^2 &=& \bigg(
M^2~(1 + \cos^2 \theta) - r_{-}~M~\cos^2 \theta - {r_{-}^2 \over 4} \sin^2 \theta \bigg)~
\bigg(
- {dt^2 + dy^2 \over y^2} + d\theta^2 \bigg) \\ \nonumber
&+&
{4 \sin^2 \theta ~M^2~ \bigg(M - {r_{-} \over 2} \bigg)^2 \over
\bigg(
M^2~(1 + \cos^2 \theta) - r_{-}~M~\cos^2 \theta - {r_{-}^2 \over 4} \sin^2 \theta \bigg)}~
\bigg( d\phi + {dt \over y} \bigg)^2.
\een
In order to put the near-horizon metric for the Kerr-Sen extremal black hole into the
Weyl-Papapetrou form, one can remind that $(y,~\theta)$ part of the line element is conformal to
$dy^2 + y^2~d\theta^2$. Thus, let us introduce the change of variables given by
\be
\rho = y ~\sin \theta, \qquad z = y~\cos \theta.
\label{coor}
\ee
Using the above relation (\ref{coor}) the line element (\ref{nh}) takes the form of (\ref{wp}).


For the near-horizon geometry it turns out that 
the fields appearing in EMAD-gravity can be determined by
\ben \label{vv}
v &=& {(2 M r_{-})^{1 \over 2}~(M - {r_{-} \over 2}) \over M^2 (1 + \cos^2 \theta) - {r^2_{-} \over 4}
~\sin^2 \theta - M~r_{- }~\cos^2 \theta}, \\ \label{kk}
\kappa &=& - {(2 M r_{-})^{1 \over 2}~(M - {r_{-} \over 2})~\cos \theta
\over  M^2 (1 + \cos^2 \theta) - {r^2_{-} \over 4}~\sin^2 \theta - M~r_{- }~\cos^2 \theta}, \\ \label{aa}
a &=& {r_{-}~\cos \theta \over (M - {r_{-} \over 2})~(1 + \cos^2 \theta)}, \\ \label{pp}
e^{2 \phi} &=& {(M - {r_{-} \over 2})~(1 + \cos^2 \theta) \over
M^2 (1 + \cos^2 \theta) - {r^2_{-} \over 4}~ \sin^2 \theta  - M~r_{- }~\cos^2 \theta},
\een
On this account, the components of the {\it current} matrix are provided by the following relations:
\be
w = {(2 M r_{-})^{1 \over 2} \over M^2 (1 + \cos^2 \theta) - {r^2_{-} \over 4}~\sin^2 \theta - M~r_{- }~\cos^2 \theta}~
\bigg[ \bigg(M - {r_{-} \over 2}\bigg)~(1 + \cos^2 \theta) - r_{-}  \bigg]~
{\cos \theta \over 1 + \cos^2 \theta},
\label{ltw}
\ee
\ben
e^{- 2 \phi}~v &=& {(2 M r_{-})^{1 \over 2} \over (M - {r_{-} \over 2})~(1 + \cos^2 \theta)}, \\
e^{- 2 \phi}~v^2 &=& {2 M r_{-} \over M^2 (1 + \cos^2 \theta) - {r^2_{-} \over 4}~\sin^2 \theta - M~r_{- }~\cos^2 \theta},
\een
and combining relations (\ref{vv}) and (\ref{ltw}) we get the following:
\be
v~w = {2 M r_{-}~(M - {r_{-} \over 2}) \over \bigg(
M^2 (1 + \cos^2 \theta) - {r^2_{-} \over 4}~\sin^2 \theta - M~r_{- }~\cos^2 \theta \bigg)^2}
~\bigg[ \bigg(M - {r_{-} \over 2}\bigg)~(1 + \cos^2 \theta) - r_{-}  \bigg]~
{\cos \theta \over 1 + \cos^2 \theta}.
\label{ltv}
\ee

\subsection{Boundary conditions}
Now we shall study the behaviour of the all components comprising the {\it deviation}
matrix $\Psi$ as one approaches near-horizon limit, axes of rotations and asymptotic spatial infinity.
Inspection of 
Eqs.(\ref{ltw}) and (\ref{ltv}) 
reveals that they are proportional to ${\cos \theta \over 1 + cos^2 \theta}$, which
can be rewritten in the form as follows
\be
{\cos \theta \over 1 + \cos^2 \theta} = {1 \over 2} - {\sin^2 \theta 
\over 4  ~(1 + \cos^2 \theta )~\cos^2 {\theta \over 2}}.
\ee
This shows that the deviation from the constant factors out through the term proportional to $\sin^2 \theta$.
By virtue of the relation described by (\ref{kk}) one gets that
\be
\kappa \simeq {\alpha \over 2} - {\sin^2 \theta \over 4~(1 + \cos^2 \theta)~\cos^2 {\theta \over 2}~
[M^2 + \cO(\sin^2 \theta)]},
\ee
where $\alpha $ is a constant. Returning to the relation (\ref{aa}) 
one can reveal that it
provides the same deviation from constant factor through term proportional to $\sin^2 \theta$. 
On the other hand, the dilaton term reduces to the following form:
\be
e^{2 \phi} \simeq {(M - {r_{-} \over 2})^2 \over M^2 + \cO (\sin^2 \theta)} \simeq \cO(1).
\label{eee}
\ee
\par
Let us consider the behaviour of the {\it deviation} matrix $\Psi$ near the event horizon of the extremal
Kerr-Sen solution given by the line element (\ref{nh}). Using Eqs.(\ref{ltw})-(\ref{eee})
one can calculate the trace of $\Psi$. It is rather tedious work but can be easily automated by using the symbolic 
math program, as in the nodegenerate case.
For $\theta \neq 0, ~\pi$ the value of $\Psi$ is finite in the limit $1/y$, when one approaches the extremal event horizon.
As we approach to the horizon for all $\theta$ including $0$ and $\pi$ the value of all the ingredients in $\Phi_{A}$
are bounded and we get $tr \Psi = \cO(1)$.
\par
Next, we shall take into consideration
the asymptotic spatial infinity. We assume that the solutions are asymptotically flat, i.e.,
the metric coefficients have the forms of (\ref{met}) and for the field under consideration relations (\ref{limit}) 
are satisfied. Inspection of all the components of the {\it deviation} matrix provides the conclusion that 
$tr \Psi = \cO(1/{\hat r}^2)$. The same behaviour was revealed in the in the non-extremal case.
\par
As far as the boundary conditions near the axis of rotation are concerned, they
are chosen to guarantee the regularity of 
the line describing extremal black hole near the fixed points of rotational symmetry described by the
adequate Killing vector field. As one approaches the extremal horizon along the axis $\theta = 0$ and
performs the necessary calculations, we
can draw the conclusion that $tr \Psi = \cO(1)$.
All these conclude the following uniqueness result:\\
{\bf Theorem}:\\
Let us consider a stationary axisymmetric spacetime with a connected degenerate future event horizon 
in four-dimensional EMAD gravity being the low-energy
limit of the heterotic string theory 
Then, any solution with the same boundary and regularity conditions as the extremal Kerr-Sen black hole
is the extremal Kerr-Sen solution itself.

\begin{acknowledgments}
This work was partially financed by the Polish budget funds in 2010 year as
the research project.
\end{acknowledgments}



\end{document}